\documentstyle[preprint,aps]{revtex}
\begin{document}
\draft
\title{Classical and Quantum Dynamics of a Periodically Driven Particle
in a Triangular Well}
\author{Michael E. Flatt\'{e}~\footnote{Electronic address:
	flatte@zeke.harvard.edu}}
\address{Division of Applied Sciences, Harvard University, Cambridge,
         Massachusetts 02138}
\author{Martin Holthaus~\footnote{Electronic address:
	holthaus@mv13a.physik.uni-marburg.de}}
\address{Fachbereich Physik der Philipps--Universit\"{a}t, Renthof 6,
	D--35032 Marburg, Germany}
\maketitle

\begin{abstract}
\baselineskip=19pt
We investigate the correspondence between classical and quantum mechanics
for periodically time dependent Hamiltonian systems, using the example of
a periodically forced particle in a one-dimensional triangular well potential.
In particular, we consider quantum mechanical Floquet states associated
with resonances in the classical phase space. When the classical motion
exhibits {\it sub}harmonic resonances, the corresponding Floquet states
maintain the driving field's periodicity through dynamical tunneling.
This principle applies both to Floquet states associated with classical
invariant vortex tubes surrounding stable, elliptic periodic orbits and to
Floquet states that are associated with unstable, hyperbolic periodic orbits.
The triangular well model also poses a yet unsolved mathematical problem,
related to perturbation theory for systems with a dense pure point spectrum.
The present approximate analytical and numerical results indicate that
quantum tunneling between different resonance zones is of crucial importance
for the question whether the driven triangular well has a dense point or an
absolutely continuous quasienergy spectrum, or whether there is a transition
from the one to the other.
\end{abstract}



\narrowtext
\vfill
\eject


\section{Introduction}

Ever since Bohr's formulation of the correspondence principle~\cite{Bohr23},
the relation between classical and quantum mechanics has been a subject of
great conceptual interest. The simplest type of time-independent, bounded
classical Hamiltonian system is integrable, i.e., if there are as
many ($n$) independent constants of motion as degrees of freedom. The phase
space of such a system is completely stratified into invariant
$n$-tori~\cite{Arnold78} and the quantum mechanical energy eigenfunctions and
eigenvalues can be determined approximately with the help of the
Einstein-Brillouin-Keller (EBK) quantization
rules~\cite{Einstein17,Brillouin26,Keller58,Percival77}.
If, on the other hand, the classical system is chaotic, then semiclassical
quantization relies on classical periodic
orbits~\cite{Gutzwiller90,CvitanovicEckhardt89,BerryKeating90,TannerEtAl91,AurichEtAl92}.
``Typical'' classical Hamiltonian systems, however, are neither completely
regular nor completely chaotic, but exhibit an intricate coexistence of
regular and apparently stochastic motion~\cite{LichtenbergLieberman83}.
An important example of this type is the three-body Coulomb
system. The application of both torus quantization and a periodic orbit cycle
expansion to this system has resulted in a semiclassical understanding of the
spectrum of the Helium atom~\cite{WintgenEtAl94}.

Much insight into the classical-quantum correspondence has been gained by
the study of simple model systems. A popular model, the periodically
kicked rotor, has led to the discovery of quantum mechanical suppression of
classical diffusion~\cite{CasatiEtAl79}, a phenomenon that could subsequently
be explained by a mechanism closely related to Anderson localization in
disordered one-dimensional crystal lattices~\cite{FishmanEtAl82,GrempelEtAl84}.
For a general quantum system defined by a periodically time dependent
Hamiltonian operator $H(t) = H(t+T)$ the role of the stationary states is
taken over by the Floquet states~\cite{Zeldovich66,Ritus66}.
If the operator
\begin{equation}
{\cal H} := H(t) - i\hbar\partial_{t} \; ,
\end{equation}
acting on an extended Hilbert space of $T$-periodic
functions~\cite{Sambe73}, has $T$-periodic eigenfunctions,
i.e., if the eigenvalue equation
\begin{equation}
[ H(t) - i\hbar\partial_{t} ] u_{\alpha}(t)
= \varepsilon_{\alpha} u_{\alpha}(t)
\label{EVE}
\end{equation}
can be solved with periodic boundary conditions in time,
$u_{\alpha}(t) = u_{\alpha}(t+T)$, then the Floquet states
\begin{equation}
\psi_{\alpha}(t) := u_{\alpha}(t)\exp(-i\varepsilon_{\alpha}t/\hbar)
\end{equation}
are solutions to the time-dependent Schr\"{o}dinger equation. The eigenvalue
equation~(\ref{EVE}) now plays a role analogous to that of the stationary
Schr\"{o}dinger equation for time-independent systems, and the objective
of a semiclassical theory is the approximate computation of the Floquet
eigenfunctions $u_{\alpha}(t)$ and the quasienergies $\varepsilon_{\alpha}$,
starting again from invariant objects in the classical phase space.

However, periodically time dependent quantum systems pose some
mathematical difficulties that are rarely met in the time-independent case.
Consider a Hamiltonian operator of the form $H(t) = H_{0} + \beta H_{1}(t)$,
where only $H_{1}(t) = H_{1}(t+T)$ carries the time dependence, and $\beta$
is a dimensionless coupling constant. Assume further that $H_{0}$ possesses
a discrete spectrum of energy eigenvalues
$E_{n}$ $(n=1,2,\ldots,\infty)$, with corresponding eigenfunctions
$\varphi_{n}$. Then for $\beta = 0$ the Floquet states can be written as
\begin{equation}
\psi_{(n,m)}(t) = \left( \varphi_{n}e^{im\omega t} \right)
\exp[-i(E_{n} + m\hbar\omega)t/\hbar]
\end{equation}
with $\omega = 2\pi/T$.
If $m$ is an integer number, the Floquet function
$u_{(n,m)}(t) = \varphi_{n}e^{im\omega t}$
is $T$-periodic, as required; the corresponding quasienergies are
$\varepsilon_{(n,m)} = E_{n} + m\hbar\omega$. The index~$\alpha$ in~(\ref{EVE})
thus becomes a double index, $\alpha = (n,m)$, and the quasienergy spectrum
is given by the energy eigenvalues {\em{modulo} $\hbar\omega$}.
Therefore it can be divided into Brillouin zones of width $\hbar\omega$.

Apart from special cases, this spectrum will be dense on the real axis.
But what happens when $\beta > 0$? Rigorous perturbation theory for such
systems with a dense point spectrum is anything but
trivial~\cite{Bellissard85,Howland92a}.
The conditions for the perturbed system to retain a pure
point spectrum constitute a crucial unsolved problem.
Partial answers to this problem
of ``quantum stability''~\cite{Howland92a} have been obtained only recently:
it has been shown by Howland~\cite{Howland89} that even for arbitrarily
large $\beta$ the quasienergy spectrum remains a dense pure point spectrum,
provided the gaps $E_{n+1} - E_{n}$ between the energy eigenvalues of $H_{0}$
grow sufficiently rapidly with $n$, and the perturbation $H_{1}(t)$ is bounded.
This result applies, for instance, to a periodically forced particle in a
one-dimensional box. The condition on the boundedness of the perturbation can
be relaxed in the case of certain other periodically forced anharmonic
oscillators~\cite{Howland92b}, but the condition on the growth of the energy
gaps between the eigenvalues of $H_{0}$ remained essential also in subsequent
studies~\cite{Nenciu93,Joye94}.

A model that violates both propositions of Howland's theorem is given by a
periodically driven particle in a triangular well potential: the unperturbed
system has a pure point spectrum, with the gaps between successive energy
eigenvalues $E_{n}$ decreasing asymptotically as $n^{-1/3}$, and its
dipole operator is unbounded. This model emerges, for example, in the
description of far-infrared irradiated semiconductor
heterojunctions~\cite{Bastard88}, or of microwave-driven charge carriers in
superfluid helium~\cite{ShimshoniSmilansky88}.

This model is the subject of the present work. Although fairly
simple, it provides interesting insights into the classical-quantum
correspondence for periodically time-dependent systems with a mixed classical
phase space. As we will demonstrate, many features of the
quantum system have a very close link to its classical counterpart,
both in the classically regular (elliptic) and chaotic (hyperbolic) case.
Nevertheless, quantum mechanical tunneling is of crucial importance to
understand prominent features of both the quasienergy spectrum and the Floquet
eigenfunctions.

The tunnel effect might also play a central role in the question
whether there is a pure point spectrum or an absolutely
continuous one.  From a rigorous point of view this difficult
question is still open,
but it has been argued that the quasienergy spectrum of the periodically
driven triangular well exhibits a transition from a point spectrum to a
continuous spectrum at a nonzero, finite value of the driving
amplitude~\cite{BenvenutoEtAl91,OliveiraGuarneriCasati94}.
Although there are well studied examples of perturbations through which
a discrete spectrum becomes continuous~\cite{FriedrichsRejto62}, a spectral
transition in a system like the driven triangular well would be quite
unusual, and therefore interesting. If there is such a transition, there
must be a mechanism producing it; however, what could that be?

Thus, the aim of this work is twofold. First, we wish to provide instructive
examples of classical-quantum correspondence and non-correspondence.
Due to its simplicity, the triangular well model might even be useful for
courses in elementary quantum mechanics, despite the open mathematical
problems. Second, we wish to stimulate further rigorous research on the
nature of the quasienergy spectrum of periodically time dependent quantum
systems. Previous
work~\cite{Bellissard85,Howland92a,Howland89,Howland92b,Nenciu93,Joye94}
has addressed this problem in a quite formal way. However, a
physically oriented starting point, such as a semiclassical analysis, might
also prove fruitful, and provide complementary information.

We proceed as follows. Section~2 introduces the classical model and states
the results of a standard resonance analysis. Unlike the periodically forced
particle in a box considered by Reichl and Lin~\cite{ReichlLin86,LinReichl88},
the triangular well model has the property that even for arbitrarily small
driving amplitude there exist arbitrarily large primary resonances in the
classical phase space. Therefore, even for small amplitudes an infinite number
of quantum mechanical Floquet states will be influenced by these resonances.
Section~3 contains a semiclassically-motivated description of near-resonant
Floquet states, followed in Section~4 by a more refined analysis, which is an
extension of previous work by Berman and Zaslavsky~\cite{BermanZaslavsky77}.
Section~5 compares the analytically calculated quasienergy spectrum with
numerical data. The central Section~6 explores the influence of classical
phase space structures on the Floquet states. We do not focus on questions
pertaining to the resonance
overlap~\cite{ReichlLin86,LinReichl88,BermanEtAl81,BermanKolovsky83},
but rather try to stretch the single-resonance picture as far as possible.
An interesting result that will be obtained in this way is the existence of
a closed link between states associated with regular classical dynamics and
others linked with apparently chaotic parts of the classical phase space:
Floquet states associated with unstable, hyperbolic periodic orbits appear
as certain excitations of states most closely associated with invariant
manifolds surrounding stable elliptic periodic orbits. The final discussion
then ties together some loose ends, and returns to the question whether or
not there could be a spectral transition.

\section{The Classical Triangular Well}

A classical particle of mass $m$ that is confined to a one-dimensional
triangular well and driven by a monochromatic force of amplitude $F$
and frequency $\omega$ is described by the Hamiltonian
\begin{equation}
\tilde{H}(P,X,t) = \frac{P^{2}}{2m} + V_{a}(X) + F X \cos(\omega t)
\end{equation}
with $V_{a}(X) = aX$ for $X \geq 0$ and $V_{a}(X) = \infty$ for $X < 0$.
Introducing dimensionless variables,
\begin{eqnarray}
x    & = & m\omega^{2}X/a	\nonumber \\
p    & = & \omega P/a		\nonumber \\
\tau & = & \omega t		\label{SIM}
\end{eqnarray}
this Hamiltonian can be written in the form
$\tilde{H}(P,X,t) = a^{2}/(m\omega^{2}) H(p,x,\tau)$,
where the scaled Hamiltonian
\begin{equation}
H(p,x,\tau) = \frac{p^{2}}{2} + V_{1}(x) + \beta x \cos(\tau)
    \equiv H_{0}(p,x) + \beta x \cos(\tau)
\label{SCH}
\end{equation}
depends only on the scaled amplitude $\beta = F/a$. A detailed
analysis of the ``kicked'' version of this system, where the sinusoidal
time dependence is replaced by a sequence of $\delta$ kicks with
alternating signs, has been presented by Shimshoni and
Smilansky~\cite{ShimshoniSmilansky88}.

To express $H(p,x,\tau)$ in terms of the action-angle variables of the
undriven system defined by $H_{0}(p,x)$, we first exploit the relation
between the action $I$ and the energy $E$:
\begin{equation}
I = \frac{1}{\pi}\int_{0}^{E}\!\sqrt{2(E - x)} dx
= \frac{1}{3\pi}(2E)^{3/2} \; .
\end{equation}
Hence,
\begin{equation}
H_{0}(I) = (3 \pi I)^{2/3}/2 \; .   \label{UPH}
\end{equation}
The oscillation frequency $\Omega(I)$
of the undriven particle in the triangular well is then found from Hamilton's
equation
\begin{equation}
\Omega(I) = \frac{\partial H_{0}}{\partial I} =
\left(\frac{\pi^{2}}{3I}\right)^{1/3} \; ,
\label{FRE}
\end{equation}
and the angle variable is $\theta = \Omega(I)\tau$, restricted to the interval
from $0$ to $2\pi$.

Imposing the initial condition $x(\tau=0) = 0$, the unperturbed trajectories
can be written as
\begin{eqnarray}
x(\theta) & = & \frac{3 I^{2/3}}{(3\pi)^{1/3}}
    \left(\theta - \frac{\theta^{2}}{2\pi}\right)	\nonumber \\
& = & \frac{I^{2/3}}{(3\pi)^{1/3}}\left(\pi - \sum_{n=1}^{\infty}
    \frac{6}{\pi n^{2}}\cos(n\theta) \right)	\; .
\end{eqnarray}
Thus, the full Hamiltonian becomes
\begin{equation}
H(I,\theta,\tau) = \frac{(3\pi I)^{2/3}}{2}
+ \frac{\beta}{3}(3\pi I)^{2/3}\cos(\tau)
- \beta\frac{(3\pi I)^{2/3}}{\pi^{2}}\sum_{n=1}^{\infty}\frac{1}{n^{2}}
\left[\cos(\tau + n\theta) + \cos(\tau - n\theta)\right] \; .
\end{equation}
Primary resonances occur when the particle's action is such that the
cycle time of unperturbed motion coincides with an integer multiple, $N$,
of the period of the driving force~\cite{LichtenbergLieberman83,Chirikov79}.
Then the argument $\tau - n\theta$ in the above expansion becomes stationary
for $n = N$. The resonant action $I_{N}$ can thus be found from the equation
$\Omega(I_{N}) = 1/N$. By~(\ref{FRE}),
\begin{equation}
I_{N} =\frac{\pi^{2}N^{3}}{3} \; .
\label{LOC}
\end{equation}

Approximating $H_{0}(I)$ quadratically around $I_{N}$ and keeping only
the resonant term of the perturbation (taken at $I = I_{N}$), one obtains the
familiar pendulum Hamiltonian~\cite{LichtenbergLieberman83,Chirikov79}
\begin{equation}
H(I,\theta,\tau) \approx H_{0}(I_{N}) + \Omega(I_{N})(I - I_{N})
+ \frac{1}{2M}(I - I_{N})^{2} - \beta\cos(\tau - N\theta) \; ,
\label{PEN}
\end{equation}
where
\begin{equation}
M^{-1} = \left.\frac{d \Omega(I)}{d I}\right|_{I_{N}}
= -\frac{1}{\pi^{2}N^{4}} \; .
\label{DFM}
\end{equation}
This Hamiltonian describes the regular, resonant motion close to $I_{N}$.
Within the pendulum approximation, the half-width of the resonances is
\begin{equation}
\Delta I_{N} = 2\pi N^{2} \sqrt{\beta} \; .
\label{WID}
\end{equation}
According to the Chirikov criterion~\cite{LichtenbergLieberman83,Chirikov79},
the $N$-th resonance overlaps with the $(N+1)$-th when
$I_{N+1} - I_{N} = \Delta I_{N+1} + \Delta I_{N}$. The corresponding
critical values of the driving strength are
\begin{equation}
\beta_{N}^{(c)} = \frac{\pi^{2}}{16}
\left(\frac{N^{2} + N + 1/3}{N^{2} + N + 1/2}\right)^{2} \; .
\label{OVR}
\end{equation}
For example, $\beta_{1}^{(c)} = 0.537$ and $\beta_{2}^{(c)} = 0.586$.

Summing up, the model of the classical driven particle in a triangular well
has the following three essential features:
(i) there are an infinite number of primary resonances, with the $N$-th
resonant action $I_{N}$ being proportional to $N^{3}$;
(ii) the half-widths $\Delta I_{N}$ grow quadratically with the order
$N$ of resonance;
(iii) within the pendulum approximation, all resonance overlaps occur in the
interval $ 0.537 < \beta < 0.617 $ of the control parameter $\beta$.

Of course, the pendulum approximation can only give a coarse description
of the actual dynamics, and is restricted to moderate values of $\beta$.
The larger $\beta$, the more important are the stochastic layers surrounding
the resonances~\cite{LichtenbergLieberman83}.
When $\beta$ is increased beyond the perturbative regime, the widths of the
zones of resonant, mainly regular motion in phase space do no longer grow
proportional to $\sqrt{\beta}$, as suggested by~(\ref{WID}), but they actually
shrink. The $\sqrt{\beta}$ growth is overcompensated by the growth of the
stochastic layers. This fact is illustrated by the following two figures.
Fig.~1 shows a Poincar\'{e} surface of section for the driven particle in a
triangular well with $\beta = 0.5$, taken for $\tau = 3\pi/2 \bmod 2\pi$.
Apart from small secondary resonances, the most prominent features are
the island of regular motion for $N =1$ in the lower left corner, and the
two islands for $N = 2$. The stochastic layers, which for smaller $\beta$
merely cover limited areas bordering the individual resonance zones,
have already merged into a connected sea of stochastic motion. The locations
of the resonant islands agree reasonably well with the predictions
of~(\ref{LOC}), but for $N=2$ the half-width is clearly smaller than
$\Delta I_{2}$, as given by~(\ref{WID}).
As shown in Fig.~2, for $\beta = 1.0$ the islands are squeezed to a small
fraction of their previous size.
But still, there exists regular motion ``inside'' the resonant islands,
although $\beta$ is significantly larger than $\beta_{\infty}^{(c)}$.

\section{Semiclassical Approach}

The scaled classical Hamiltonian~(\ref{SCH}) depends only on a single
dimensionless parameter. A second one appears in quantum mechanics:
Planck's constant $\hbar$ is scaled to
\begin{equation}
\hbar_{eff} := m\omega^{3}\hbar/a^{2}
\end{equation}
as a consequence of the transformations~(\ref{SIM}).
The model of the periodically driven triangular well, therefore, is of
paradigmatic simplicity for a study of the classical-quantum correspondence:
one parameter, $\beta$, controlls the degree of nonintegrability in the
classical system, the other, $\hbar_{eff}$, controlls the scale at which
its quantum mechanical counterpart can resolve classical phase space
structures.

Since $I - I_{N}$ and $\theta$ constitute a canonically conjugate pair, the
approximate pendulum Hamiltonian~(\ref{PEN}) can be quantized by the
replacement
\begin{equation}
I - I_{N} \longrightarrow \frac{\hbar_{eff}}{i}\frac{\partial}{\partial\theta}
\label{REP}
\end{equation}
This is essentially a semiclassical argument, since it ignores the fact that
the canonical transformation $(p,x) \rightarrow (I,\theta)$ does not commute
with the quantization operation. The resulting Schr\"{o}dinger equation
\begin{equation}
i\hbar_{eff}\frac{\partial}{\partial \tau} \psi(\theta,\tau) =
\left[ H_{0}(I_{N})
+ \Omega(I_{N})\frac{\hbar_{eff}}{i}\frac{\partial}{\partial\theta}
- \frac{\hbar_{eff}^{2}}{2M}\frac{\partial^{2}}{\partial\theta^{2}}
- \beta\cos(N\theta - \tau) \right] \psi(\theta,\tau)
\label{SGL}
\end{equation}
can be solved with the help of the ansatz
\begin{equation}
\psi(\theta,\tau) \equiv \chi(z)\exp(-iW\tau/\hbar_{eff}) \; ,
\label{ANS}
\end{equation}
where the new variable $z$ is given by
\begin{equation}
z = (N\theta - \tau)/2 \; .
\end{equation}
Inserting~(\ref{ANS}) into (\ref{SGL}) and using the resonance condition
$\Omega(I_{N}) = 1/N$, one obtains the Mathieu
equation~\cite{AbramowitzStegun72}
\begin{equation}
\left( \frac{d^{2}}{dz^{2}} + a - 2q\cos(2z) \right) \chi(z) = 0
\label{MAT}
\end{equation}
with parameters
\begin{eqnarray}
a & = & \frac{8|M|}{\hbar_{eff}^{2}N^{2}}(H_{0}(I_{N}) - W)  \label{SCA} \\
q & = & \frac{4|M|}{\hbar_{eff}^{2}N^{2}}\beta  \; .	\label{SCQ}
\end{eqnarray}
The general solution to this equation has the Floquet form
$\chi(z) = P_{\nu}(z)e^{i\nu z}$, with a characteristic exponent $\nu$ and
a $\pi$-periodic function $P_{\nu}(z) = P_{\nu}(z+\pi)$. Thus, the
corresponding wave functions $\psi(\theta,\tau)$ have the form
\begin{equation}
\psi(\theta,\tau) = P_{\nu}\left(\frac{N\theta - \tau}{2}\right)
\exp\left(i\nu\frac{N\theta - \tau}{2} - i\frac{W\tau}{\hbar_{eff}}\right) \; .
\label{THE}
\end{equation}
Since $\theta$ is an angle variable, $\psi$ must be $2\pi$-periodic
in $\theta$. This requirement now restricts the possible values of $\nu$
to the discrete set
\begin{equation}
\nu = \nu(j) = \frac{2j}{N} \;\;\; , \;\;\; j = 0,1,2,\ldots,N-1 \; .
\label{NUC}
\end{equation}
According to~(\ref{SCQ}), the parameter $q$ entering the Mathieu equation is
fixed by the choice of the amplitude $\beta$. The allowed values $\nu(j)$ can
occur as characteristic exponents of solutions to the Mathieu equation only
for certain values (labelled by the integer $k$) $a_{k}(\nu(j),q)$ of the
other parameter $a$. These values $a_{k}(\nu(j),q)$ finally determine,
by~(\ref{SCA}), the allowed $W_{j,k}$. The wave
functions~(\ref{THE}) are now fully specified. Since they already are of the
required form $\psi_{j,k}(\theta,\tau) =
u_{j,k}(\theta,\tau)\exp(-i\varepsilon_{j,k}\tau/\hbar_{eff})$,
with $u_{j,k}(\theta,\tau) = u_{j,k}(\theta,\tau + 2\pi)$, the quasienergies
for the approximate equation~(\ref{SGL}) have been found:
\begin{eqnarray}
\varepsilon_{j,k} & = & \left[\hbar_{eff}\frac{\nu(j)}{2} + W_{j,k}\right] \;
\bmod \hbar_{eff}
\nonumber \\
& = & \left[H_{0}(I_{N}) - \frac{\hbar_{eff}^{2}N^{2}}{8|M|}a_{k}(\nu(j),q)
+ \hbar_{eff}\frac{j}{N} \right]\; \bmod \hbar_{eff} \; .
\label{SCS}
\end{eqnarray}

The cases $N = 1$ and $N = 2$ are of particular interest.
For $N = 1$ there is $j = 0$ only, and the $a_{k}(0,q)$ coincide with the
well known characteristic values~\cite{AbramowitzStegun72} that give rise to
$\pi$-periodic Mathieu functions. Those values associated with even Mathieu
functions are customarily denoted~\cite{AbramowitzStegun72}
as $ a_{0}(q), a_{2}(q), a_{4}(q), \ldots $, and those associated with
odd functions as $ b_{2}(q), b_{4}(q), b_{6}(q), \ldots$ With the definition
\begin{equation}
\alpha_{0,k}(q) := \left\{ \begin{array}{rcl}
    a_{k}(q)   & , & k = 0,2,4,\ldots \\
    b_{k+1}(q) & , & k = 1,3,5,\ldots \end{array}\right.
\label{RE1}
\end{equation}
the approximate quasienergies for an $N=1$ resonance can be written as
\begin{equation}
\varepsilon_{k} = \left[H_{0}(I_{N})
- \frac{\hbar_{eff}^{2}}{8|M|}\alpha_{0,k}(q) \; \bmod \hbar_{eff} \right]\; .
\label{SP1}
\end{equation}

For $N=2$ there are two groups of states, labelled by $j=0$ and $j=1$.
Those with $j =1$ have to be constructed from the $2\pi$-periodic Mathieu
functions. The required characteristic values
$\alpha_{1,k}(q) \equiv a_{k}(1,q)$ are
\begin{equation}
\alpha_{1,k}(q) := \left\{ \begin{array}{rcl}
    b_{k+1}(q) & , & k = 0,2,4,\ldots \\
    a_{k}(q)   & , & k = 1,3,5,\ldots \end{array}\right. \; .
\end{equation}
Thus, the quasienergies are given by
\begin{equation}
\varepsilon_{j,k} = \left[H_{0}(I_{N})
- \frac{\hbar_{eff}^{2}}{2|M|}\alpha_{j,k}(q)
+ \hbar_{eff}\frac{j}{2} \right]\; \bmod \hbar_{eff}    \label{SP2}
\end{equation}
for $j = 0,1$ and $k = 0,1,2,\ldots$

This spectrum has a simple interpretation~\cite{HolthausFlatte94}.
The Mathieu equation~(\ref{MAT}) can be regarded as a stationary
Schr\"{o}dinger equation for a fictitious particle moving in a cosine
lattice, and the solution for the $N$-th resonance requires periodic boundary
conditions after $N$ cosine wells.
Thus, $N=1$ leads to a single-well potential, and $N=2$ to a double well.
The {\em{quasi-}}energies of the original problem correspond to the
{\em{energies}} of the fictitious particle.
The eigenstates in the single well are simply labelled by the quantum
number $k$. The spectrum of the double well, however, consists of a sequence
of doublets, each one associated with a symmetric and an antisymmetric
eigenfunction. The absolute value of the difference
$\varepsilon_{1,k} - \varepsilon_{0,k} -\hbar_{eff}/2 =:
(\Delta\varepsilon)_{k}$
of eigenvalues within each doublet corresponds to the tunnel splitting.
The well known asymptotic behavior of the characteristic
values~\cite{AbramowitzStegun72},
\begin{equation}
b_{k+1}(q) - a_{k}(q) \sim 2^{4k+5}\sqrt{\frac{2}{\pi}}
\frac{q^{k/2+3/4}}{k!}\exp(-4\sqrt{q})	\; ,
\label{DAB}
\end{equation}
shows that the tunnel splitting becomes exponentially small with $\sqrt{q}$,
i.e., with the square root of the driving amplitude $\beta$ (see~(\ref{SCQ})).
That is easy to understand: $q$ determines the depth of the individual cosine
wells; the larger $q$, the weaker the tunneling between them.
In particular, the {\em{ground state splitting}}, i.e., the (quasi-) energy
difference between the two states most tightly bound by the double well
potential, is
\begin{eqnarray}
(\Delta\varepsilon)_{0}
& = & \frac{\hbar_{eff}^{2}}{2|M|}\left(b_{1}(q) - a_{0}(q)\right) \nonumber \\
& \sim & \frac{\hbar_{eff}^{2}}{\pi^{2}}\sqrt{\frac{2}{\pi}}q^{3/4}
    \exp(-4\sqrt{q})
\label{TUS}
\end{eqnarray}
with $q = 16\pi^{2}\beta/\hbar_{eff}^{2}$. It is quite remarkable that
the approximate construction of near-resonant Floquet states for
$N > 1$ accounts for the tunnel effect, even within the present semiclassical
approach.

\section{Quantum Mechanical Approach}

The considerations in the previous section, albeit instructive, have several
flaws. Most of all, it is not easy to control the accuracy of the
semiclassical replacement~(\ref{REP}), and an expression like $H_{0}(I_{N})$
will, in general, not correspond directly to a quantum mechanical energy
eigenvalue.
It is, therefore, desirable to analyze the role of classical resonances
in quantum systems in a strictly quantum mechanical way, without direct
recourse to classical mechanics. In this section, we sketch such a quantum
mechanical approach. It is basically a generalization of the previous work by
Berman and Zaslavsky~\cite{BermanZaslavsky77}. Although it also confirms the
main conclusions of the semiclassical computation, its primary merit lies in
the fact that it clearly shows the physical and mathematical difficulties that
will be encountered in the attempt to improve the approximation.

We consider the Hamiltonian~(\ref{SCH}) and denote the unperturbed energy
eigenstates and eigenvalues of the triangular well by $|n\rangle$ and $E_{n}$,
respectively. We assume that the $r$-th eigenstate is ``resonant'', i.e.,
that the level spacing close to the $r$-th eigenstate is approximately
equal to $\hbar_{eff}/N$ for some integer $N$:
\begin{equation}
E'_{r} \equiv \left.\frac{d E_{n}}{d n}\right|_{n=r}
\approx \frac{\hbar_{eff}}{N}	\; .
\end{equation}
This is the quantum mechanical equivalent of the classical resonance
condition $\Omega(I_{N}) = 1/N$.
Only for special choices of $\hbar_{eff}$ will it be possible to obtain
an exact identity $E'_{r} = \hbar_{eff}/N$; in the general case we choose $r$
such that the absolute value of the difference
\begin{equation}
\delta := E'_{r} - \frac{\hbar_{eff}}{N}    \label{DEL}
\end{equation}
is minimized. At most, $\delta$ will be of the order of $E''_{r}$.
We also stipulate that $r$ be so large that for levels close to $E_{r}$
the spacing between $E_{n}$ and $E_{n+N}$ varies only slowly on the scale of
$\hbar_{eff}$, which implies the inequality
\begin{equation}
E''_{r} \ll E'_{r}  \; . \label{UNG}
\end{equation}

We are interested in a solution $|\psi(\tau)\rangle$ to
the time-dependent Schr\"{o}dinger equation that consists mainly of a
superposition of near-resonant eigenstates. Thus, a reasonable ansatz is
provided by
\begin{equation}
| \psi(\tau) \rangle = \sum_{n} c_{n}(\tau) | n \rangle
\exp\left\{-i\left( E_{r} + (n-r)\frac{\hbar_{eff}}{N}\right)\frac{\tau}
{\hbar_{eff}}\right\} \; ,
\end{equation}
which immediately leads to
\begin{equation}
i\hbar_{eff}\dot{c}_{n}(\tau) = \left(E_{n} - E_{r}
- \hbar_{eff}\frac{n-r}{N}\right)c_{n}(\tau)
+ \beta\cos(\tau)\sum_{m} \langle n | x | m \rangle c_{m}(\tau)
\exp[i(n-m)\tau/N] \; .
\label{EQC}
\end{equation}
Keeping in mind the inequality~(\ref{UNG}), we expand the eigenvalues
quadratically around $E_{r}$,
\begin{equation}
E_{n} \approx E_{r} + (n-r)E'_{r} + \frac{1}{2}(n-r)^{2}E''_{r} \; .
\label{QUA}
\end{equation}

According to our propositions, the time-dependence of the coefficients
$c_{n}(\tau)$ should be weak, compared to that of the exponentials
$\exp(i\ell\tau/N)$ with a nonzero integer $\ell$. Correspondingly, we keep
only the stationary terms $m = n \pm N$ on the r.h.s.\ of~(\ref{EQC}).
Thus each coefficient $c_{n}(\tau)$ is coupled only to $c_{n \pm N}(\tau)$,
$c_{n \pm 2N}(\tau)$, and so forth. One obtains $N$ disjoint sets of
coefficients, which we denote by ${c_{n}^{(j)}}$.
Within each set, $j = 0,1,\ldots, N-1$ is constant, and $n$ is restricted to
the values $n = r + j + mN$, where $m$ is an integer.
In addition, we assume that the dipole matrix elements
$\beta\langle n | x | n \pm N \rangle$ can be approximated by a constant
$V$, independent of $n$. Using now the new index $m$ to label the
coefficients, one arrives at the $N$ decoupled sets of equations:
\begin{equation}
i\hbar_{eff}\dot{c}_{m}^{(j)}(\tau) =
\frac{1}{2} N^{2}E''_{r}(m + j/N)^{2}c_{m}^{(j)}(\tau)
+ N\delta(m + j/N)c_{m}^{(j)}(\tau)
+ \frac{V}{2}\left(c_{m+1}^{(j)}(\tau) + c_{m-1}^{(j)}(\tau)\right) \; .
\label{NDC}
\end{equation}
To solve these equations, we set
\begin{eqnarray}
c_{m}^{(j)}(\tau) & = & \frac{1}{2\pi}\int_{0}^{2\pi}\! d\varphi \,
f_{j}(\varphi) e^{-im\varphi} e^{-iW_{j}\tau/\hbar_{eff}} \nonumber \\
& = & \frac{1}{2N\pi}\int_{0}^{2N\pi}\! d\varphi \,
g_{j}(\varphi) e^{-i(m+j/N)\varphi} e^{-iW_{j}\tau/\hbar_{eff}} \; ,
\end{eqnarray}
with $W_{j}$ still to be determined.
Assuming that $f_{j}(\varphi)$ is $2\pi$-periodic, the auxiliary function
$ g_{j}(\varphi) \equiv f_{j}(\varphi)\exp(ij\varphi/N)$ is $2N\pi$-periodic.
The resulting equation for $g_{j}(\varphi)$, i.e.,
\begin{equation}
\left(-\frac{1}{2}N^{2}E''_{r} \frac{d^{2}}{d\varphi^{2}}
+ \frac{N\delta}{i} \frac{d}{d\varphi} + V \cos(\varphi) - W_{j} \right)
g_{j}(\varphi) = 0\;,
\end{equation}
can then be transformed into a standard Mathieu equation; the ansatz
\begin{equation}
g_{j}(\varphi) = \exp[-2iz\delta/(NE''_{r})]\chi_{j}(z)
\end{equation}
with $\varphi = 2z$ leads to
\begin{equation}
\left(\frac{d^{2}}{dz^{2}} + a -2q\cos(2z) \right) \chi_{j}(z) = 0 \; ,
\label{MAQ}
\end{equation}
with the parameters
\begin{eqnarray}
a & = & \frac{8W}{N^{2}E''_{r}} + \left(\frac{2\delta}{NE''_{r}}\right)^{2}
\label{QMA} \\
q & = & \frac{4V}{N^{2}E''_{r}} \; .
\label{QMQ}
\end{eqnarray}
Starting from a general Floquet solution to the Mathieu equation~(\ref{MAQ}),
$\chi_{j}(z) = P_{\nu(j)}(z)\exp(i\nu(j) z)$, the function $g_{j}(\varphi)$
can now be expressed as
\begin{equation}
g_{j}(\varphi) = P_{\nu(j)}(\varphi/2)
\exp\left(i\frac{\nu(j)\varphi}{2} - i\frac{\varphi\delta}{NE''_{r}}\right) \;
{}.
\end{equation}
Since, on the other hand,
$g_{j}(\varphi) = f_{j}(\varphi)\exp(ij\varphi/N)$,
one finds $f_{j}(\varphi) = P_{\nu(j)}(\varphi/2)$, together with
\begin{equation}
\nu(j) = \frac{2j}{N} + \frac{2\delta}{NE''_{r}} \;\; ; \;\;
j = 1, \ldots \, , N-1 \; .
\label{NUQ}
\end{equation}
As in the previous section, these characteristic exponents admit only a
discrete set of allowed values for the parameter $a$; these values are again
denoted by $a_{k}(\nu(j),q)$.
By~(\ref{QMA}), they lead to a discrete set of possible values $W_{j,k}$
for the quantities $W_{j}$. Now the expansion coefficients are completely
specified:
\begin{eqnarray}
c_{m}^{(j)}(\tau) & = & \frac{1}{2\pi}\int_{0}^{2\pi} \! d\varphi \,
P_{\nu(j),k}(\varphi/2)\exp(-im\varphi -iW_{j,k}\tau/\hbar_{eff}) \nonumber \\
& \equiv & \chi_{j,k}^{(m)}\exp(-iW_{j,k}\tau/\hbar_{eff}) \; .
\end{eqnarray}
Thus, the resulting approximate Floquet solutions $| \psi_{j,k}(\tau) \rangle$
to the Schr\"{o}dinger equation can finally be written as
\begin{eqnarray}
|\psi_{j,k}(\tau)\rangle & = & \sum_{m} \chi_{j,k}^{(m)}
| r + j + mN \rangle e^{-im\tau}
\exp\left\{-i\left( E_{r} + W_{j,k} + \hbar_{eff}\frac{j}{N}\right)
\frac{\tau}{\hbar_{eff}}\right\}    \nonumber \\
& \equiv & | u_{j,k}(\tau) \rangle \exp(-i\varepsilon_{j,k}\tau/\hbar_{eff})
\; . \label{AFL}
\end{eqnarray}
Using~(\ref{QMA}), the quasienergies
$\varepsilon_{j,k} = E_{r} + W_{j,k} + \hbar_{eff}j/N$ ($\bmod \, \hbar_{eff}$)
are
\begin{equation}
\varepsilon_{j,k} = \left[E_{r} + \frac{1}{8} N^{2}E''_{r}a_{k}(\nu(j),q)
-\frac{\delta^{2}}{2E''_{r}} + \hbar_{eff}\frac{j}{N}
\right]\; \bmod \hbar_{eff} \; .
\label{QMS}
\end{equation}
The structure of the near-resonant quasienergy spectrum obtained in this
way is equal to the structure of~(\ref{SCS}), but the parameters differ.
These differences are small in the regime where the approximations
are valid. First, the exact energy eigenvalues for the triangular well
potential are given by
\begin{equation}
E_{n} = \left(\frac{\hbar_{eff}^{2}}{2}\right)^{1/3}\! z_{n} \; ,
\end{equation}
where the numbers $z_{n}$ ($n=1,2,3,\ldots$) denote the negative zeros of the
Airy function $Ai(z)$~\cite{AbramowitzStegun72}:
\begin{equation}
z_{n} = f\left(\frac{3\pi}{2}(n-\frac{1}{4})\right)
\end{equation}
with
\begin{equation}
f(z) \sim z^{2/3}\left(1 + \frac{5}{48}z^{-2} - \frac{5}{36}z^{-4}
+ \ldots \right) \; .
\end{equation}
If this expansion is approximated by its leading term, and if the Maslov
correction resulting from the right turning point of the motion in the
triangular well potential is neglected, one has
\begin{equation}
z_{n} \approx \left(\frac{3\pi}{2}(n-\frac{1}{4})\right)^{2/3}
\approx (3\pi n/2)^{2/3}    \; ,
\end{equation}
and hence
\begin{equation}
E_{n} \approx \frac{1}{2}(3\pi\hbar_{eff}n)^{2/3}   \; .
\end{equation}
Since the quantum number $r$ of the resonant state is large by assumption,
these ``semiclassical'' approximations are appropriate.
Thus, within these approximations the eigenvalues $E_{n}$ agree with those
obtained from the classical Hamiltonian $H_{0}(I)$ by the Bohr-Sommerfeld
replacement $I \rightarrow \hbar_{eff}n$, see~(\ref{UPH}).
With $I_{N} \approx \hbar_{eff}r$, the parameter $M^{-1}$ in~(\ref{SCS})
approximates $E''_{r}/\hbar_{eff}^{2}$ in~(\ref{QMS}), see~(\ref{DFM}).

The dipole matrix elements of the triangular well can be computed
analytically~\cite{Gordon69}:
\begin{equation}
\langle n | x | m \rangle = \left\{ \begin{array}{ll}
    (2\hbar_{eff})^{2/3}z_{n}/3              &, \; n = m    \\
   -(2\hbar_{eff})^{2/3}/(z_{n} - z_{m})^{2} &, \; n \neq m
		\end{array} \right. \; .
\end{equation}
Using the arithmetic mean
\begin{equation}
V = \frac{\beta}{2}\left( \langle r | x | r + N \rangle
		   + \langle r | x | r - N \rangle \right)
\end{equation}
for the coupling constant $V$ appearing in~(\ref{NDC}), and expanding
$z_{r\pm N}$ to first order in $N/r$, one finds
\begin{equation}
V \approx -\beta \; .
\end{equation}
Then the semiclassical and the quantum expression for the Mathieu
parameter~$q$, (\ref{SCQ}) and~(\ref{QMQ}), coincide.

The characteristic exponents~(\ref{NUC}) are exactly equal to those
in~(\ref{NUQ}) only if $\delta$ vanishes, i.e., if $E'_{r}$ is exactly
equal to $\hbar_{eff}/N$. In the general case, there can therefore be a slight
difference between the values of the parameters $a_{k}(\nu(j),q)$ in both
approximation schemes. However, since these parameters lie between
the characteristic values $a_{k}(q)$ and $b_{k+1}(q)$, and since
$b_{k+1}(q) - a_{k}(q)$ vanishes for large $q$ according to~(\ref{DAB}),
this difference is visible only for small $q$. The term
$-\delta^{2}/(2E''_{r})$ appearing in~(\ref{QMS}) makes sure that the
approximate quasienergies merge into the quadratically approximated
eigenvalues~(\ref{QUA}) for vanishing $q$. Since $\delta$ is, at most,
of the order of $E''_{r}$, one sees from~(\ref{UNG}) and~(\ref{DEL})
that this term is small compared to $\hbar_{eff}/N$.
Thus, there is a one-to-one correspondence between the semiclassically
calculated spectrum~(\ref{SCS}) and~(\ref{QMS}).

The approximations made in the quantum calculation have their obvious
analogues in the previous section. There is one detail that deserves
particular attention: it has tacitly been assumed that the Fourier
coefficients $\chi_{j,k}^{(m)}$ of the approximate Floquet states~(\ref{AFL})
decay so rapidly that the non-existing ``eigenstates below the ground
state $|1\rangle$'' carry zero weight. The approximations leading to the
Mathieu equation introduce an additional symmetry (the translational
invariance of the cosine lattice) which renders the problem integrable.
The larger $q$, the broader the distribution of the Fourier coefficients.
When the distribution becomes so broad that the wave function can feel
deviations from the lattice symmetry, the Mathieu approximation becomes
insufficient.

\section{The Quasienergy Spectrum: Analytical versus Numerical Results}

To compute quasienergies and Floquet states for the periodically forced
triangular well~(\ref{SCH}) numerically, we expand the wave functions in a
truncated basis of unperturbed eigenfunctions
$\langle x | n \rangle = \varphi_{n}(x)$, with $n = 1, \ldots\, n_{max}$.
Within this basis set, we compute the monodromy operator (i.e., the
``one-cycle evolution operator'') $U(2\pi,0)$. Its eigenvalues are the
Floquet multipliers $\exp(-i2\pi\varepsilon_{n}/\hbar_{eff})$;
the eigenvectors yield the Floquet functions $u_{n}(x,0)$ at time $\tau = 0$.
The numerical stability of the results obtained in this way has been checked by
varying the truncation border $n_{max}$. However, even though the results
appear stable, it is by no means certain whether such a numerical procedure
can, for sufficiently large basis size, actually ``converge'' in the strict
mathematical sense. This question is related to a deeper physical problem
which we shall discuss in the final section; for the moment, we take the
numerical results at face value.

The following two figures show parts of the quasienergy spectrum versus the
driving amplitude $\beta$. Because of the $\bmod \, \hbar_{eff}$ structure
of the spectrum, each of the (infinitely many) Floquet states has a
representative quasienergy within the first quasienergy Brillouin zone
$-\hbar_{eff}/2 \leq \varepsilon < +\hbar_{eff}/2$. That brings about the
necessity to select from all computed quasienergies only a small fraction
for graphical display. The computed spectra contain a plethora of avoided
crossings, and it is not always possible to select the quasienergies in such
a way that the displayed eigenvalues appear to vary smoothly with $\beta$.
Therefore, abruptly ending or beginning lines visible in the figures are not
caused by some sort of numerical instability; such discontinuities are merely
unavoidable in the selection procedure.

First, we choose $\hbar_{eff} = 1.33325$. With this choice, one has
$E'_{20} = \hbar_{eff}/2$, so that the energy eigenstate $r=20$ is
exactly resonant for $N=2$. Fig.~3 shows {\em{only}} quasienergies
originating from states close to the resonant one.  From~(\ref{SP2})
one expects two almost identical groups of eigenvalues,
separated by $\hbar_{eff}/2$. This expectation is fully confirmed by the
numerical result; quasienergies belonging to one of the two groups
are labelled by the Mathieu quantum numbers $k$ in the right margin of Fig.~3.
In view of the corresponding classical mechanics, the fact that for
$\beta = 0.5$ {\em{at least}} $2 \times 8$ states are accessible to the
Mathieu approximation is quite remarkable: the upper right corner
of the phase space portrait displayed in Fig.~1 shows the area
$2\pi\hbar_{eff}$, corresponding to the area ``occupied'' by a single Floquet
state. Judging by the total areas of the islands of regular motion,
one might expect the classical $N=2$ resonance to influence roughly
$2 \times 4$ Floquet states, merely half the actual number. However, it
must be borne in mind that the actual regime of influence of a classical
resonance is not restricted to the regular islands. In their vicinity
there are still lots of fragments of invariant manifolds, such as Cantori.

Fig.~4 displays quasienergies for $\hbar_{eff} = 0.66219$, so that
$r=40$ is exactly resonant for $N=2$, and the number of states accessible
to the Mathieu approximation is doubled. To check the accuracy of the
previous analytical considerations, Fig.~5 shows the evaluation
of~(\ref{QMS}) for the same parameters as used in Fig.~4, and $k=0, \ldots, 9$.
The semiclassical expression $M^{-1}$ here differs by merely $0.047\%$
from $E''_{40}/\hbar_{eff}^{2}$; the Mathieu parameter $q$ is approximately
given by $16\pi^{2}\beta/\hbar_{eff}^{2}$. The characteristic values
$a_{k}(\nu(j),q)$ have been computed by an algorithm suggested
in~\cite{LindnerFreese94}, which has been found to be both efficient and
accurate. The agreement between Figs.~4 and~5 is quite good, although
one of the approximations used in the derivation of~(\ref{QMS}) is
certainly critical: the assumption that the matrix elements
$\langle n | x | n \pm 2 \rangle$ close to $n = r = 40$ can simply be
approximated by a constant is investigated in Fig.~6. Although these
matrix elements vary substantially in the range from $n=30$ to $50$, the
comparison of Figs.~4 and~5 indicates that the somewhat crude approximation
is still viable. This observation is rendered plausible by a perturbative
argument: a better approximation to the ``nearest neighbor couplings''
in~(\ref{NDC}) would be given by a linear fit,
$\beta\langle n | x | n \pm N \rangle \approx V + (n-r)V'$,
leading to a perturbation
\begin{equation}
H_{pert} \approx \frac{V'}{2}\sum_{n}(n-r) \left( |n\rangle \langle n+N|
     + |n\rangle \langle n-N| \right) \; .
\end{equation}
The matrix elements
$\langle u_{j,k}(\tau) | H_{pert} | u_{j,k}(\tau) \rangle$,
taken with the approximate Floquet states~(\ref{AFL}), then contain only
contributions proportional to either $e^{-i\tau}$ or to $e^{+i\tau}$.
Therefore, to first order in $H_{pert}$ the approximate
quasienergies~(\ref{QMS}) remain unchanged:
\begin{equation}
\int_{0}^{2\pi} \! d\tau \,
\langle u_{j,k}(\tau) | H_{pert} | u_{j,k}(\tau) \rangle = 0 \; .
\end{equation}
However, the significance of the Rayleigh-Schr\"{o}dinger perturbation series
for operators acting in the extended Hilbert space of $2\pi$-periodic
functions~\cite{Sambe73} is questionable. This is again a problem with
perturbing of a dense point spectrum; the $\bmod \, \hbar_{eff}$ structure
gives rise to an infinite number of small denominators in the expression
for the second order perturbative contribution of $H_{pert}$.

\section{Classical-Quantum Correspondence}

Figs.~7 and 8 show contour plots of the probability density
$ |u_{j,k}(x,\tau)|^{2} $ of a (numerically computed) member of the ground
state doublet $(k=0)$ for the $N=2$ resonance and $\beta = 0.5$, with
$\hbar_{eff} = 1.33325$ and $0.66219$, respectively. In these and the
following figures, the spatial coordinate $x$ ranges from $x_{min} = 0$
to $x_{max} = 25$, and the displayed interval of time is $8\pi$,
corresponding to 4~cycles of the driving force.

The structure of these Floquet states, as well as that of the quasienergy
spectrum of near-resonant states, can intuitively be understood if the
quantum dynamics is connected in a direct way to the corresponding
classical mechanics.
Whereas quantum mechanical energy eigenstates of systems with a (partially)
integrable classical counterpart are associated with invariant manifolds in
the usual phase space spanned by momentum and position
coordinates~\cite{MaslovFedoriuk81}, Floquet states of a (partially)
integrable system with a Hamiltonian that is $2\pi$-periodic in $\tau$ are
associated with invariant, $2\pi$-periodic vortex tubes~\cite{Arnold78}
in the odd-dimensional, extended phase space
$\{(p,x,\tau)\}$~\cite{BreuerHolthaus91}.
Consequently, there are semiclassical quantization rules for Floquet states
that resemble closely the well-known Einstein-Brillouin-Keller
rules~\cite{Einstein17,Brillouin26,Keller58,Percival77} for
energy eigenstates. In the present case of a system with merely one spatial
degree of freedom and an effective Planck constant $\hbar_{eff}$, these rules
can be formulated as follows:
First, the correct ``quantized'' vortex tubes are selected by
\begin{equation}
\oint_{\gamma_{1}} \! p dx = 2\pi\hbar_{eff}\left(k + \frac{1}{2}\right) \; ,
\label{RU1}
\end{equation}
where the quantization path $\gamma_{1}$ winds once around a $2\pi$-periodic
tube in a plane $\tau = const.$, and $k = 0,1,2,\ldots$ is an integer.
The quasienergies then follow from the second rule
\begin{equation}
\varepsilon = -\frac{1}{2\pi}\int_{\gamma_{2}} (pdx - Hd\tau)
+ m\hbar_{eff} \; ,
\label{RU2}
\end{equation}
where the path $\gamma_{2}$ is $2\pi$-periodic in $\tau$ and lies on such a
tube, with the integration extending over one period. The integer $m$
accounts for the $\bmod \, \hbar_{eff}$ structure of the quasienergy
spectrum~\cite{BreuerHolthaus91}.

It is now crucial to realize that the direct computation of Floquet
states by these rules is possible only if there actually are vortex tubes
that inherit the $2\pi$-periodicity of the underlying Hamiltonian. The rules
can, therefore, be applied to vortex tubes that are merely perturbative
(and, hence, $2\pi$-periodic) deformations of energy manifolds of the
undriven system, or to the vortex tubes of an $N = 1$ resonance.
Such a resonance emerges if the time of one cycle of unperturbed motion is
equal to the driving period, $2\pi$. Hence the central elliptic periodic
orbit of such a resonance is also $2\pi$-periodic, and the rules~(\ref{RU1}),
(\ref{RU2}) refer to the vortex tubes surrounding it. The fact that these
tubes are not simply deformations of energy manifolds of the undriven system
is reflected by the emergence of a new quantum number, $k$. According
to (\ref{RU1}), the resonant ground state, $k=0$, is associated with the
innermost quantized tube surrounding the elliptic periodic orbit; the first
excited state, $k=1$, with the next largest quantized tube, and so forth.
Naively, one might expect the hierarchy of resonant states to extend
as far as there are still preserved tubes.

Obviously, the quantum number $k$ that appears in the semiclassical
rule~(\ref{RU1}) can be identified with the Mathieu quantum number $k$
in~(\ref{SP1}); for example, the innermost quantized vortex tube corresponds
to the state that is most tightly bound in the cosine potential.
In the example shown in Fig.~1, the closed curves seen inside the
$N=1$ island in the lower left corner are sections of $2\pi$-periodic vortex
tubes with the plane $\tau = 3\pi/2$, and~(\ref{RU1}) applies, provided that
$\hbar_{eff}$ is small enough so that the required tubes actually
``fit inside'' the resonance.

An $N=2$ resonance, however, requires a different
reasoning~\cite{HolthausFlatte94}. In this case, the central elliptic
periodic orbits, as well as the surrounding vortex tubes, are $4\pi$-periodic.
Of course, one can still use~(\ref{RU1}) to select ``quantized'' vortex tubes
and compute the associated semiclassical wave functions. In principle, such
a procedure can be carried through in some detail~\cite{MirbachKorsch94}, but
the wave functions constructed in this way are not Floquet states. Generally,
the association of a wave function with an invariant classical object
living in the extended phase space is witnessed by the concentration of
the probability density along the projection of that object to the
``configuration space'' $\{(x,\tau)\}$. Hence, wave functions associated
with a $4\pi$-periodic vortex tube would also be $4\pi$-periodic in time,
whereas a Floquet state necessarily has to be $2\pi$-periodic.

This difficulty can only be resolved if a Floquet state is not associated
with a {\em{single}}, $4\pi$-periodic tube, but rather with {\em{both parts}}
that result from its projection to the fundamental piece
$\{(p,x,\tau) , 0 \leq \tau \leq 2\pi\}$ of extended phase
space~\cite{HolthausFlatte94}. In this way, $2\pi$-periodicity can be
restored. The existence of two equivalent families of classical
$4\pi$-periodic tubes, displaced from each other in time by a single period,
imply that the associated quantum mechanical Floquet states must appear in
pairs. That corresponds precisely to the Mathieu approximation for $N=2$; the
eigenfunctions of the double cosine well are not localized in the individual
wells (as the corresponding Floquet states of the original problem are not
associated with a single invariant vortex tube), but are equally extended
over both wells (as the Floquet states are equally associated with both tubes),
so that doublets of states emerge, separated in (quasi-) energy by the tunnel
splitting.

The additional quasienergy separation of $\hbar_{eff}/2$ between the two
groups of states with $j=0$ and $j=1$ in~(\ref{SP2}) has a simple explanation
in terms of an analogue borrowed from solid state physics.
If the lattice constant of a one-dimensional lattice is doubled from
$d$ to $2d$, the width of the crystal momentum Brillouin zone shrinks from
$2\pi/d$ to $\pi/d$. A similar ``dimerization'' occurs here:
in the vicinity of an $N=2$ resonance the classical invariant manifolds are
$2\cdot 2\pi$-periodic, so that the width of the corresponding quasienergy
Brillouin zone should be $\hbar_{eff}/2$, instead of $\hbar_{eff}$. Thus,
the separation by $\hbar_{eff}/2$ can be regarded as the manifestation of a
purely classical effect. In contrast, the additional tunnel splitting is of
genuinely quantum mechanical origin.

The ``ground state'' Floquet states displayed in Figs.~7 and~8 likewise
both reflect the dynamics of the corresponding classical system and exhibit
a purely quantum mechanical effect. A particle following the stable elliptic
periodic orbit of the $N=2$ resonance bounces against the wall of the
triangular well, is reflected, runs against the linear slope, is decelerated,
reverses its direction and hits the wall again precisely $4\pi$ after the
previous bounce. There is an equivalent periodic orbit displaced in time
from the first one by $2\pi$. The two quantized vortex tubes with $k=0$
surround these orbits in the extended phase space, and, as seen in Figs.~7
and~8, the quantum mechanical probability density of an associated Floquet
state is concentrated along the projection of {\em{both}} tubes to the
$(x,\tau)$ plane. Now the projection of a tube that describes classical
particles moving away from the wall intersects, at certain moments, the
projection of a tube describing motion towards the wall. In the intersection
regions there is a strong interference of both quantum mechanical
``possibilities''. The two equivalent classical tubes are isolated in the
extended phase space, but the associated wave functions communicate.

If one forms even or odd linear combinations of the eigenfunctions of a
symmetric double well, one obtains functions that are localized in the
individual wells. In complete analogy, it is possible to form superpositions
of the two members of the resonant ground state doublet that are localized
along the projection of only one of its two tubes. This is demonstrated in
Fig.~9, again for $\hbar_{eff} = 0.66219$ and $\beta = 0.5$. The wave function
displayed there appears to follow only one of the two equivalent tubes and,
thus, manifestly exhibits subharmonic motion. However, it remains coupled
to the other tube by quantum tunneling: after the tunneling time
$ \tau_{tunnel} := \hbar_{eff}\pi/(\Delta\varepsilon)_{0} $
the density would be concentrated along the projection of the other tube,
and then start to tunnel back. The approximation~(\ref{TUS}) yields
$ (\Delta\varepsilon)_{0} \approx 8.5 \cdot 10^{-24} $,
and thus a tunneling time
$ \tau_{tunnel} \approx 4 \cdot 10^{22} \cdot 2\pi $.

The same principle holds for the excited resonant states,
i.e., for those states with higher Mathieu quantum number $k$. For example,
Fig.~10 shows the density of a pure Floquet state with $k = 6$, for the
same parameters as before. According to~(\ref{RU1}), its associated tubes are
much wider than those for $k = 0$. Hence, the projections are larger than
those of the ground state tubes, which is clearly reflected by the wave
functions. Again, it is possible to form a superposition that is localized
along only one of the tubes (see Fig.~11), but now the tunneling time is
$11$ orders of magnitude shorter than that for $k=0$.

The vortex tubes for $k=6$, $\hbar_{eff} = 0.66219$ and $\beta = 0.5$
still roughly ``fit inside'' the $N=2$ islands of regular motion, see Fig.~1.
But the quasienergy spectrum displayed in Fig.~4
indicates that the number of states accessible
to the Mathieu approximation is {\em{larger}} than the area of the regular
islands, divided by $2\pi\hbar_{eff}$; the hierarchy of near-resonant states
extends farther than one might expect. Further evidence for this statement
is provided by the following two figures. Fig.~12 shows the probability
density of a Floquet state with $k = 10$, whereas Fig.~13 shows the density
of a superposition with its partner state. The tubes for $k=10$ do not fit
into the islands of regular motion; they would lie in the surrounding
stochastic sea. Nevertheless, the wave functions are structured similarly
to those in Figs.~10, 11, as if there still were ``quantizable'' vortex tubes.
Hence, although fully preserved vortex tubes do no longer exist outside the
islands, it {\em{must}} be possible to use their remnants, such as Cantori,
in a semiclassical quantization procedure. In this way, the quantum mechanical
regularity associated with regular classical resonant motion is continued
into a regime where a major part of the corresponding classical dynamics
is chaotic. Expressed differently, the Mathieu quantum number $k$
is not necessarily related to invariant vortex tubes; the number of Floquet
states that can be labelled by $k$ is larger than the number of vortex tubes
that fit into the regular islands. Even though the transition from (mainly)
regular to (mainly) stochastic classical motion appears to be quite sharp on
the borders of the islands of stability, there is no correspondingly sharp
transition in quantum mechanics.

It had already been pointed out that the Mathieu equation~(\ref{MAT}) can be
regarded as the Schr\"{o}dinger equation for a fictitious particle in a cosine
lattice. When the energy of such a particle is close to the top of the barrier,
i.e., when $a_{k}(\nu(j),q) \approx 2q$, its probability density is strongly
enhanced in the vicinity of the potential maxima, reflecting the fact that a
corresponding classical particle would spend most of its time there.
Just as the minima of the cosine potential correspond, within the pendulum
approximation, to the stable, elliptic periodic orbits, the maxima correspond
to the unstable, hyperbolic ones. According to this qualitative consideration,
Floquet states with a Mathieu quantum number $k$ such that
$a_{k}(\nu(j),q) \approx 2q$ should be separatrix states, with a probability
density that is strongly enhanced along the unstable, hyperbolic periodic
orbits.

Quantitatively, the simple Mathieu approximation becomes less accurate for
large quantum numbers $k$. For the present parameters, a characteristic value
with $k=17$ is closest to $2q$, whereas it is actually a state with $k=16$
whose density exhibits the clearest traces of the unstable classical orbit
(see Fig.~14). Nervertheless, the approximation still yields the correct
order of magnitude for the tunnel splitting: analytically, one obtains
$ | (\Delta\varepsilon)_{16}/\hbar_{eff} | \approx 1.4 \cdot 10^{-2}$,
compared to the numerically found value of $ 4.6 \cdot 10^{-2} $.
Naturally, close to the top of the cosine wells the splitting is
quite large (in the present example, the absolute value of
$(\Delta\varepsilon)_{16}$ is $21$ orders of magnitude larger than the
ground state splitting!), and the probability
density of both members of a doublet shows significant differences.
In spite of that, Fig.~15 demonstrates that it is still possible to form
a superposition of a separatrix state with its partner state such that,
at least for a short time, the resulting density is strongly enhanced along
only one of the two unstable $4\pi$-periodic orbits. But complete
destructive interference along the other one is no longer possible.
The large tunnel splitting results in a quite short tunneling time
for this doublet, $\tau_{tunnel} \approx 10.9 \cdot 2\pi$. Thus, after
merely 11 cycles of the driving force the density will be concentrated along
the other hyperbolic orbit.

Quantum mechanical energy eigenstates with a density that peaks along
unstable, hyperbolic periodic orbits, first investigated by McDonald
and Kaufmann~\cite{McDonaldKaufmann88} and Heller~\cite{Heller84},
have been dubbed ``scars''. In the context of periodically driven systems,
such scars have attracted some interest since it was speculated that
the experimentally observed anomalous stability of certain hydrogenic
Rydberg states against microwave ionization~\cite{Koch92} could be related
to a scarred Floquet state~\cite{JensenEtAl89}. In fact, the scar investigated
by Jensen et al.~\cite{JensenEtAl89} turns out to be simply the separatrix
state of an $N=1$ resonance.
The present simple model shows, first of all, why an {\em{unstable}}
classical orbit can ``attract'' quantum mechanical probability density,
and it also shows under which conditions this phenomenon can occur in
Floquet states. To get further insight into a possible connection between
scarring and enhanced stability against ionization, it might be interesting
to investigate also an $N=2$ resonance in highly excited, microwave-driven
hydrogen atoms: are there, again, anomalously stable states?

\section{Implications}

The preceding investigation of Floquet states and quasienergies linked to
a primary $N=2$ resonance has demonstrated that the incompatibility of the
$2\pi$-periodic boundary conditions imposed on the quantum mechanical Floquet
states, and the $2 \cdot 2\pi$ periodicity of classical invariant manifolds,
necessarily leads to the tunnel effect.
Pictorially speaking, there is quantum tunneling from one of the $N=2$ islands
seen in Fig.~1 through the stochastic sea to the other. Thus, one is faced
with a special type of ``dynamical tunneling'', as studied by Davis and
Heller~\cite{DavisHeller86} in the context of energy eigenstates.
The generalization to the case of an arbitrary $N$ is obvious: the Mathieu
approximation leads to a cosine lattice with periodic boundary conditions
after $N$ wells. The wave functions associated with classical,
$N\cdot 2\pi$-periodic vortex tubes correspond to the single-well ``atomic''
states of that lattice, whereas the actual near-resonant Floquet states
correspond to delocalized Bloch waves.

Besides dynamical tunneling, another feature known from the investigation
of time-independent systems carries over to the periodically time-dependent
case. Jaff\'{e}, Shirts and Reinhardt~\cite{JaffeReinhardt82,ShirtsReinhardt82}
have argued that remnants of destroyed invariant manifolds, so-called ``vague
tori'' existing in the apparently chaotic part of phase space, may introduce
enough regularity to allow EBK-like quantization even in the absence of
complete tori. Analogously, ``vague vortex tubes'' should be responsible for
the surprisingly regular appearance of the Floquet state displayed in Fig.~12.
The notion of a vague torus, although somewhat hard to formulate in precise
mathematical terms, provides a bridge between regular and fully chaotic
classical dynamics, and gives some insight how a quantum system behaves
when its classical counterpart falls into this transition regime.
The numerical example of the previous section also shows that there can be
a closed link, expressed by the existence of a common quantum number $k$,
between Floquet states associated with invariant manifolds surrounding stable,
elliptic periodic orbits and states associated with unstable, hyperbolic
periodic orbits. In the general case, one might wish to apply torus
quantization to the classically regular, elliptic motion and a periodic orbit
cycle expansion to the chaotic, homoclinic motion~\cite{WintgenEtAl94}.
But to which extent, if at all, is it possible to ``interpolate'' between both
quantization schemes? One-dimensional, periodically time-dependent systems
like the one studied here, with their merely three-dimensional phase
space $\{(p,x,\tau)\}$, might provide the simplest access to this important
problem.

Returning now to the specific properties of the triangular well model,
it has been suggested that there exists a transition from
a pure point quasienergy spectrum to a continuous one at a nonzero
value of the amplitude~$\beta$~\cite{BenvenutoEtAl91,OliveiraGuarneriCasati94}.
A point spectrum would be connected with square-integrable Floquet
states that have a ``localized'' distribution of coefficients when expanded
in the basis of unperturbed triangular well eigenstates; a continuous spectrum
would be connected with ``extended'' states. The classical dynamics indeed
shows a qualitative
change when $\beta$ is enlarged: for small amplitudes, all trajectories are
either tied to invariant vortex tubes or move in the stochastic layers
surrounding the individual resonance zones, but remain bounded. Within a
comparatively small interval of $\beta$, {\em all} resonances overlap
(see~(\ref{OVR})), so that the individual stochastic layers merge into a
connected stochastic sea. We then have a dichotomy: the trajectories are
either confined to the remnants of the regular resonant islands or they move
in the stochastic sea (cf.~Fig.~2), where they can gain an unlimited amount
of energy. It would be tempting to speculate that the latter type of behavior
is linked to ``extended'' Floquet states in the quantum system.

The numerical evidence remains inconclusive. Fig.~16 shows the near-resonant
quasienergy spectrum ($N=2$) for $\hbar_{eff} = 0.66219$, for amplitudes
$\beta$ between~$0$ and~$1.0$. With increasing $\beta$, more and more
resonant vortex tubes are destroyed (cf.~Figs.~1 and~2), and the number
of states that can be described by the Mathieu approximation is diminished.
Nevertheless, there are clear signs of the resonance-induced regularity
even for $\beta = 1.0$. Fig.~17 shows the probability density of a member
of the resonant ground state doublet for this amplitude, whereas Fig.~18
shows the density of a ``one-tube'' superposition with its partner state.
Although these wave functions appear considerably more ``ragged'' than
they did for $\beta = 0.5$ (cf.~Figs.~8 and~9), there does not seem to be
a sharp transition, but merely a gradual change of behavior. However, the
quantum mechanics of the periodically driven triangular well is more subtle
than a numerical investigation might suggest.

The previous analytical deliberations were concerned with single resonances.
But for a rigorous study of the nature of the quasienergy spectrum the
single-resonance approximation is insufficient. To get a glimpse of the
problem, consider the approximate Mathieu spectrum~(\ref{QMS}) for the
$N$-th resonance, together with that for the $(N+1)$-th, projected into
one Brillouin zone. At certain values of $\beta$, approximate eigenvalues
belonging to the $N$-th resonance will cross others of the $(N+1)$-th.
However, the full Hamiltonian of the driven triangular well has no symmetry
which could allow these crossings: the terms neglected in the derivation of
the two near-resonant spectra will lead to a small, but nonvanishing coupling
of the Floquet states belonging to the two different resonances. Hence, the
eigenvalues will repel each other~\cite{NeumannWigner29}, so that the
crossing of the approximate eigenvalues become anticrossings, most of them
much too tiny to be detectable by numerical means. Accordingly, the Floquet
states become linear combinations of the approximate single-resonance states.
In other words, there is quantum tunneling not only between invariant
manifolds belonging to the same $N$, but also between those belonging to
different $N$. Now there are not only two primary resonances, but infinitely
many. According to~(\ref{WID}), the width $\Delta I_{N}$ of the classical
resonance zones grows quadratically with $N$, so that even for arbitrarily
small $\beta$ there will always be an infinite number of primary resonances
that support an arbitrarily large number of Floquet states, no matter how
large (or small) the effective Planck constant $\hbar_{eff}$. If one imagines
all the infinitely many near-resonant quasienergy spectra superimposed in
a single Brillouin zone for, say, $0 \leq \beta \leq 1$, with all crossings
replaced by anticrossings, one gets an extremely complicated net, with
arbitrarily narrow anticrossings appearing everywhere. It seems unavoidable
to conclude that the true quasienergies of the driven triangular well can
not be differentiable with respect to the amplitude $\beta$.

A question that could shed some light on the nature of the quasienergy
spectrum can therefore be phrased as follows: if one starts from the
approximate near-resonant Floquet states, and then takes into account quantum
tunneling between different resonances, are the resulting Floquet states
``localized'' over a finite number of resonances, or ``extended''?
It appears quite possible that, because of tunneling, there is no spectral
transition except for $\beta = 0$, and all Floquet states might be extended
for every $\beta > 0$. Or could localized and extended states coexist, such
that the number of extended states gradually increases when $\beta$ is
enlarged?

For a physicist, the approximate analytical and numerical results provide
a description of the dynamics over time scales that are short compared
to inter-resonance tunneling times, which may be sufficient for most
practical purposes. From the point of view of a rigorous mathematician,
the present study of the periodically driven triangular well certainly poses
additional questions. It suggests, however, a more direct
starting point for the investigation of periodically time dependent quantum
systems than the one adopted
in~\cite{Bellissard85,Howland92a,Howland89,Howland92b,Nenciu93,Joye94}.
The model of the kicked rotor~\cite{CasatiEtAl79}, which was motivated by
a single-resonance approximation, has meanwhile been understood in some
detail~\cite{FishmanEtAl82,GrempelEtAl84,Bellissard85,Howland92a}.
The next generation of problems seems to arise from the interaction of
different resonances; a detailed analysis of the role of quantum tunneling
is required. The triangular well
model~\cite{ShimshoniSmilansky88,BenvenutoEtAl91,OliveiraGuarneriCasati94}
appears to be ideally suited for that purpose.

\begin{figure}
\caption[FIG1]{Poincar\'e surface of section for the driven particle in a
triangular well potential~(\ref{SCH}), taken at $\tau = 3\pi/2 \bmod 2\pi$.
The island in the lower left corner originates from the $N=1$ resonance,
the two large islands from the resonance with $N=2$.
The driving amplitude is $\beta = 0.5$, lower than $\beta_{1}^{(c)} = 0.537$
which is the value required by the Chirikov criterion for the overlap
of these two resonances. The two boxes in the upper right corner enclose
areas $2\pi\hbar_{eff}$, with $\hbar_{eff} = 1.33325$ and $0.66219$,
respectively.}
\end{figure}

\begin{figure}
\caption[FIG2]{Poincar\'e section for $\beta = 1.0$. The driving amplitude
now is significantly higher than $\beta_{\infty}^{(c)} = 0.617$, above which,
according to the Chirikov criterion, all resonances overlap.
Nevertheless, there are still remnants  of the regular islands for $N=1$ and
$N=2$. The two boxes in the upper right corner have the same meaning as in
Fig.~1.}
\end{figure}

\begin{figure}
\caption[FIG3]{Part of the numerically computed quasienergy spectrum for
$\hbar_{eff} = 1.33325$, restricted to one Brillouin zone.
Energies of unperturbed triangular well eigenstates close to the ``resonant''
state $r=20$ differ by approximately $\hbar_{eff}/2$.
The behavior of the quasienergies of near-resonant states is well described
by the Mathieu approximation~(\ref{SP2}); integers in the right margin are the
quantum numbers $k$ for one of the two resonant ladders.}
\end{figure}

\begin{figure}
\caption[FIG4]{Part of the numerically computed quasienergy spectrum for
$\hbar_{eff} = 0.66219$. For this value of $\hbar_{eff}$, energies of
unperturbed triangular well eigenstates close to the ``resonant'' state
$r=40$ differ by approximately $\hbar_{eff}/2$.}
\end{figure}

\begin{figure}
\caption[FIG5]{Approximate quasienergy spectrum according to the Mathieu
approximation~(\ref{QMS}), for the same parameters as used in Fig.~4,
and $k = 0, \ldots, 9$.}
\end{figure}

\begin{figure}
\caption[FIG6]{Negative dipole matrix elements $\langle n | x | n+2 \rangle$
in the vicinity of $n=40$. Within the Mathieu approximation leading to
Fig.~5, it is assumed that all these elements are equal to unity.}
\end{figure}

\begin{figure}
\caption[FIG7]{Contour plot of the probability density of a member of the
ground state doublet ($k=0$), for $\beta = 0.5$ and $\hbar_{eff} = 1.33325$.
The density is concentrated along the two classically equivalent ``innermost
quantized vortex tubes'' surrounding the two central elliptic periodic orbits
of the $N=2$ resonance, projected to the $(x,\tau)$ plane.
Strong interference occurs where both projections intersect.
In this and the following figures, the displayed interval of space
ranges from $x_{min} = 0$ to $x_{max} = 25$, the interval of time
corresponds to $4\cdot 2\pi$, i.e., to $4$~cycles of the driving force.}
\end{figure}

\begin{figure}
\caption[FIG8]{Probability density of a member of the ground state doublet
($k=0$), for $\beta = 0.5$ and $\hbar_{eff} = 0.66219$. The wave function is
now tied closer to the associated classical object than it is in the case
of the larger $\hbar_{eff}$, see Fig.~7.}
\end{figure}

\begin{figure}
\caption[FIG9]{Probability density of a solution to the time-dependent
Schr\"odinger equation that consists of a superposition of both members of the
ground state doublet. The parameters are again $\beta = 0.5$,
$\hbar_{eff} = 0.66219$.
The interval of time corresponds to 4 cycles of the driving force. The wave
function appears to follow only one of the two quantized vortex tubes, and
thus exhibits subharmonic motion. However, it is coupled to the other tube by
the tunnel effect. The tunneling time is
$\tau_{tunnel} \approx 4 \cdot 10^{22} \cdot 2\pi$.}
\end{figure}

\begin{figure}
\caption[FIG10]{Probability density of a member of the doublet with $k=6$.
The associated vortex tubes still roughly fit inside the regular elliptic
islands, cf.~Fig.~1 ($\beta = 0.5$, $\hbar_{eff} = 0.66219$).}
\end{figure}

\begin{figure}
\caption[FIG11]{Solution to the time-dependent Schr\"{o}dinger equation that
consists of a ``one-tube'' superposition of the $k=6$ doublet
($\beta = 0.5$, $\hbar_{eff} = 0.66219$).}
\end{figure}

\begin{figure}
\caption[FIG12]{Probability density of a member of the doublet with $k=10$.
Although the associated vortex tubes would not fit inside the regular
elliptic islands seen in Fig.~1, this eigenfunction is structured
analogously to that displayed in Fig.~10
($\beta = 0.5$, $\hbar_{eff} = 0.66219$).}
\end{figure}

\begin{figure}
\caption[FIG13]{Solution to the time-dependent Schr\"{o}dinger equation that
consists of a superposition of both members of the $k=10$ doublet
($\beta = 0.5$, $\hbar_{eff} = 0.66219$).}
\end{figure}

\begin{figure}
\caption[FIG14]{Probability density of a member of the $k=16$ doublet.
This is a separatrix state: its density is concentrated along the projections
of the two hyperbolic periodic orbits originating from the $N=2$ resonance
($\beta = 0.5$, $\hbar_{eff} = 0.66219$). The arrows on the bottom indicate
the moments where the stable, elliptic periodic orbits hit the wall,
cf.~Fig.~7.}
\end{figure}

\begin{figure}
\caption[FIG15]{Solution to the time-dependent Schr\"{o}dinger equation that
consists of both members of the $k=16$ doublet. This wave function shows
signs of ``scarring'', i.e., its density appears strongly enhanced along the
projection of a single hyperbolic periodic orbit of the $N=2$ resonance.
However, the tunneling time for this doublet is merely $10.9 \cdot 2\pi$,
so that the density will be enhanced along the other hyperbolic periodic
orbit after merely $11$ cycles
($\beta = 0.5$, $\hbar_{eff} = 0.66219$).}
\end{figure}

\begin{figure}
\caption[FIG16]{Part of the quasienergy spectrum for $\hbar_{eff} = 0.66219$,
and amplitudes $\beta$ ranging from~$0$ to~$1.0$. Even for the highest
amplitude, there are clear signs of the resonance-induced regularity.}
\end{figure}

\begin{figure}
\caption[FIG17]{Probability density of a Floquet state with $k=0$ for
$\hbar_{eff} = 0.66219$ and $\beta = 1.0$, cf.~Fig.~2.}
\end{figure}

\begin{figure}
\caption[FIG18]{Superposition of both members of the $N=2$ ground state doublet
for $\hbar_{eff} = 0.66219$ and $\beta = 1.0$.}
\end{figure}

\end{document}